\documentclass[manuscript]{aastex}
\usepackage{natbib}
\usepackage{graphicx}
\usepackage{txfonts}
\shorttitle{The INTEGRAL high energy cut-off distribution of type 1 AGN}
\shortauthors{Malizia et al.}

\begin{document}


\title{The INTEGRAL high energy cut-off distribution of type 1 AGN}


\author{A. Malizia,  M. Molina, L. Bassani, J. B. Stephen}
\affil{IASF-Bologna, INAF, Via Gobetti 101, 40129 Bologna, Italy  }

\and

\author{A. Bazzano, P. Ubertini}
\affil{IAPS-Roma, INAF, Via Fosso del Cavaliere 100, I-00133 Roma, Italy}

\and
\author{A. J. Bird}
\affil{School of Physics and Astronomy, University of Southampton, SO17 1BJ, Southampton, U.K.}

\email{malizia@iasfbo.inaf.it}


\begin{abstract}
In this letter we present the primary continuum parameters, the photon index $\Gamma$ and the high energy cut-off E$_{\rm c}$, of 41 type-1 Seyfert galaxies extracted from the \emph{INTEGRAL} complete sample of AGN.
We performed a broad band (0.3-100 keV) spectral analysis by fitting simultaneously the soft and hard X-ray spectra obtained by \emph{XMM} and \emph{INTEGRAL/IBIS}-\emph{Swift/BAT} respectively in order to investigate  the general properties of these parameters in particular their distribution and mean values. We find a mean photon index for the whole sample of 1.73 with a standard deviation of 0.17 and a mean high energy cut-off of 128 keV with a standard deviation of 46 keV.
This is the first time that the cut-off energy is constrained in a such large number of AGN. 
We have 26 measurements of the cut-off, which corresponds to 63\% of the entire sample, distributed between 50 and 200 keV. There are a further 11 lower limits mostly below 300\,keV. 
Using the main parameters of the primary continuum,  we have been able to obtain the actual physical parameters of the Comptonizing  region i.e. the plasma temperature 
kT$_{\rm e}$ from 20 to 100  keV and the optical depth  $\tau$$<$4. 
Finally, with the high S/N spectra starting to come from \emph{NuSTAR} it will soon be possible to better constrain the cut-off values in many AGN, allowing the determination of more physical 
models and so to better understand the continuum emission and geometry of the region surrounding black holes.
\end{abstract}


\keywords{galaxies: active --- gamma rays: galaxies --- X-rays: galaxies}



\section{Introduction}

The advent of high energy observatories like \emph{INTEGRAL} and \emph{Swift} has provided greater depth
to the study of active galactic nuclei (AGN), as they provide information on spectral features which cannot be explored 
without observations performed above 10\,keV.
The high energy data are crucial not only to estimate  the slope of the continuum emission over a wide energy band but also
to measure the high energy cut-off and the reflection fraction, which  are important physical parameters.

Firstly, they are important parameters for evaluating the AGN contribution to the cosmic X-ray background (CXB); indeed, while
the fraction of the resolved  CXB into discrete sources is 100\% determined below 2-10\,keV,
50\% in the 7-10\,keV band, it becomes negligible at higher energies i.e. above 10\,keV, just where its spectral intensity peaks at around 30\,keV.
In order to reproduce the shape of the CXB, synthesis models (e.g. \citealt{comastri:2005}) use 
several parameters such as the fraction of  heavily obscured sources, the so-called Compton thick AGN characterized by N$_{\rm H}\geq$10$^{24}$\,cm$^{-2}$,
the coverage and the geometry of the cold gas distributed around the black hole responsible for the reflection hump,
and the high energy cut-off of the primary continuum emission,  which is not  well determined.
Recently, for the estimation of the AGN contribution to the CXB,
several models have been proposed  \citep{gandhi:2007, gilli:2007, comastri:2004} which take into account 
the average power-law photon index (and its standard deviation in values) and the spread of the average cut-off energy 
instead of the single mean values.
Therefore, the  determination of photon indices  and cut-off energies, their mean values and their distributions  over a wide sample of sources, 
covering  a wide range of energies (above 100\,keV), 
is  essential  to obtain a much  firmer estimation of the AGN contribution to the CXB at high energy.

Measuring  the slope of the continuum emission  and the high energy cut-off of AGN is also important  
because both parameters  enable us to understand the physical characteristics and the geometry of the region around the central nucleus. 
AGN spectral models that  focus  on the reproduction of  the observed shape of the primary continuum ascribe the
power-law to the inverse Compton scattering of soft photons off 
hot electrons located above the accretion disk in the so-called corona \citep{maraschi:1997, zdziarski:1998}.
In the framework of this disk-corona system, the temperature kT$_{\rm e}$ and the optical depth $\tau$ of the scattering electrons mainly determine the spectral slope, 
while the cut-off energy is related essentially to kT$_{\rm e}$, thus  simultaneous measurements of $\Gamma$
and E$_{c}$ allow us to understand  the physical parameters of the Comptonizing region.
The more  accurate  the measurements on these parameters, the better we can determine  the 
geometry and the physical properties of the inner region of the AGN.

Up to now, several studies have been carried out to specify  the distribution of the photon indices 
both in the soft 2--10\,keV \citep{bianchi:2009} and hard 20--100\,keV X-ray bands \citep{molina:2013} while, 
after early results coming in the 1990s from  the \emph{BeppoSAX}  which had a broad spectral coverage (2--100\,keV),  
measurements of high energy cut-offs have been limited by the scarcity of observations above 10\,keV.

With the launch of \emph{INTEGRAL} and \emph{Swift} the number of AGN detected above 10\,keV has grown enormously
but to properly measure the high energy cut-off, a broad-band spectral study is needed.
This implies that both low and high energy spectra have to be fitted simultaneously employing spectra with high statistical quality
such as those acquired, for example, with \emph{XMM-Newton} and \emph{INTEGRAL}.
It has been demonstrated \citep{molina:2009, derosa:2012, panessa:2011} that the match between 
\emph{XMM} and \emph{INTEGRAL} spectra is  good with a cross-calibration constant between 
the two instruments around 1. Therefore, when  a constant different from unity is found, this generally indicates  flux variations between the soft and the hard X-ray domain.\\

In this letter we report on the  high energy cut-off measurements and their distribution,
derived from the analysis of \emph{XMM} plus \emph{INTEGRAL/IBIS} and \emph{Swift/BAT} data 
of all 41 Seyfert 1 galaxies out of the 88 sources listed in the \emph{INTEGRAL} complete sample of AGN \citep{malizia:2009}.

\section{Broad Band Spectral Analysis}
In this work we have performed 0.3-100 keV spectral analysis of all type 1 Seyfert galaxies 
included in the \emph{INTEGRAL} complete sample of AGN, excluding only  5 narrow line Seyfert 1 galaxies which were extensively studied by \citet{panessa:2011}.
For all 41 AGN we have collected \emph{XMM-Newton} observations in the soft band and 
\emph{INTEGRAL/IBIS} plus \emph{Swift/BAT} observations in the hard X-ray domain.

EPIC-pn \citet{turner:2001} data were processed using the \emph{XMM-Newton} Standard
Analysis Software (SAS) version 12.0.1 employing the latest available calibration files
and following the spectral data reduction as in \citet{molina:2009}.
Since our sources are bright in X-rays, several of them were affected by pile-up and
for them only patterns corresponding to single events (PATTERN = 0) were considered.
The \emph{INTEGRAL} data reported here consist of ISGRI data from several pointings  
between revolutions 12 and 530, i.e. those coming from the fourth IBIS catalogue \citep{bird:2010}.
ISGRI data analysis and the average source spectra extraction has been obtained following the
procedure  described by  \citet{molina:2013}.
The Swift/BAT spectra \citep{baumgartner:2013}, taken in order to improve the statistics at high energies, 
are from the latest 70-month catalogue\footnote{http://swift.gsfc.nasa.gov/results/bs70mon/}. 
IBIS and BAT analysis has been performed in the 20-100 keV and 14-100 keV band, respectively. 

It is worth noting that the broad-band spectral analysis of a large fraction of our objects has 
been previously reported in \citet{molina:2009}, but \emph{XMM} and/or \emph{Swift/BAT} data were available at that time only for a limited number of AGN and the rest had a significantly poorer spectral quality at low energies.
For 3 sources (NGC 3783, IC 4329A and IGR J21247-5058) we do not perform a new spectral fitting and
use the values in \citet{molina:2009}. This is due to the fact that: a) we already had good constraints
on the spectral shape and b) the data have been recently confirmed by 
\emph{Suzaku} (NGC 3783, \citealt{brenneman:2011, patrick:2011}; IGR J21247-5058 \citealt{tazaki:2010}) and \emph{NuSTAR} 
(IC 4329A \citealt{brenneman:2013, harrison:2013}). \\

We fitted simultaneously the soft and the hard X-ray spectra obtained by \emph{XMM} and \emph{INTEGRAL/IBIS}-\emph{Swift/BAT}
in the 0.3--100\,keV energy range,
using \texttt{XSPEC} v.12.8.0, with errors quoted at 90\% confidence level for one parameter of interest ($\Delta\chi^{2}$ = 2.71).
To fit the 0.3--100\,keV continuum, we considered a  baseline model commonly used to fit broad-band X-ray 
spectra, composed of an exponentially cut-off power-law reflected from neutral matter (\texttt{PEXRAV} model in \texttt{XSPEC}), 
where R, defined as $\Omega$/2$\pi$, is the parameter which corresponds to the reflection component.
Since we are analyzing type 1 sources, we have assumed a face-on geometry and fixed  the inclination angle of the reprocessor  
to 30$^{\circ}$.

At soft energies Galactic absorption has always 
been considered and, when required by the data, intrinsic absorption in terms of simple or/and
complex, cold or ionized absorbers have been added (\texttt{WABS, PCFABS, ZXIPCF}). Since the low energy part
of type 1 AGN spectra often shows clear signs of a soft excess, this has been generally fitted with 
a thermal component (\texttt{BREMSS}) when present. A gaussian component has also been included, to take into account the  
presence of the iron K$\alpha$ line at around 6.4\,keV; in a few cases, residuals around 7\,keV have also
been fitted adding a second gaussian line to take into account the iron K$\beta$ feature.
Finally, to take into consideration possible flux variations between instruments, we have introduced 
cross-calibration constants which we left free to vary, keeping in mind that possible mis-calibration 
between \emph{XMM} and \emph{IBIS} or \emph{BAT} could mimic or hide the Compton reflection 
component above 10\,keV \citep{derosa:2012}.
In the cases of NGC 4151, NGC 4593 and 4C 74.26, the fit has been performed over a more
restricted energy range (2--100\,keV), in order to avoid complications due to low energy features, 
such as the complex warm absorber, whose treatment is beyond the scope of this work.  
Here we focus on the primary continuum parameters, i.e. photon index and cut-off energy and to properly interpret these results 
we also report the low energy components (soft-excess and absorption).
For completeness  we  list calibration constants and goodness of fit  (see  Table 1). 
Since the primary aim of the present letter is to study the cut-off measurements and to determine their distribution, 
we will not report the whole spectral parameters, such as  reflection fraction and iron line features,
and defer to an upcoming work the discussion on the spectral complexity
of each type 1 AGN of the \emph{INTEGRAL} complete sample.

As shown in the table, there are 6 sources where, in order to have a good broad-band fit, it has been 
necessary to fix the photon index to the best fit value found using the \emph{XMM} data alone;
in five of them both the values of the \emph{BAT} and \emph{IBIS} cross-calibration constants 
are consistent with unity, indicating that the continuum shape at energies above 20\,keV is consistent with that in the
2--10\,keV band. Instead in LEDA 168563 the a cross-calibration constant between \emph{XMM} and \emph{BAT} is in the range [0.53--0.65],
probably due to variability status of the source at high energy as the difference between the \emph{BAT} and \emph{IBIS} 20--100\,keV fluxes
suggests (F$_{\rm BAT}$ = 3.8$\times$ 10$^{-11}$erg\,cm$^{-2}$\,s$^{-1}$ and
F$_{\rm IBIS}$ = 5.3$\times$ 10$^{-11}$erg\,cm$^{-2}$\,s$^{-1}$).
For B3 0309+411B and IGR J18027-1455 we have used only the \emph{BAT} data for the 20--100\,keV band, 
since the former source has been detected by \emph{IBIS} in a burst map (see \citealt{bird:2010}) and the latter
has been found to be extremely variable (see \citealt{molina:2009}); in both cases, the extrapolation 
of the high energy spectrum from the 2--10\,keV band gave a source 
status more consistent with the \emph{BAT} measurement and hence we chose  not to use the \emph{IBIS} data. 
Only for IGR J18249-3243 is there no \emph{BAT} observation available; moreover, this source has also been
detected by \emph{IBIS} by means of the burst analysis, so the overall 
\emph{ISGRI} spectrum, extracted from the 4$^{\rm th}$ \emph{IBIS} catalogue is of very poor quality ($\sim$ 3$\sigma$). 
Therefore for this source the spectral parameter values should be taken with some caution.
Care should also be used for IGR J16119-6039, as this galaxy is inside the Norma cluster (A3627) 
and thus, as already pointed out by Ajello et al (2010), we should expect contamination from the cluster thermal emission
in the high energy spectrum of this source.

\section{Results}
In Table 1 we report the fit results only for the spectral parameters which are of interest in this work; the goodness of our fits (see last column in the table)
is expressed in terms of reduced $\chi^{2}$ and, as a further indication, we also report the values of the two cross-calibration constants 
between \emph{XMM} and \emph{BAT}  and between \emph{XMM} and \emph{IBIS}, which are generally both close to unity,
with the exception of a few variable sources (e.g. NGC 4151, MCG+08-11-011). 

\begin{figure}
\epsscale{.80}
\plotone{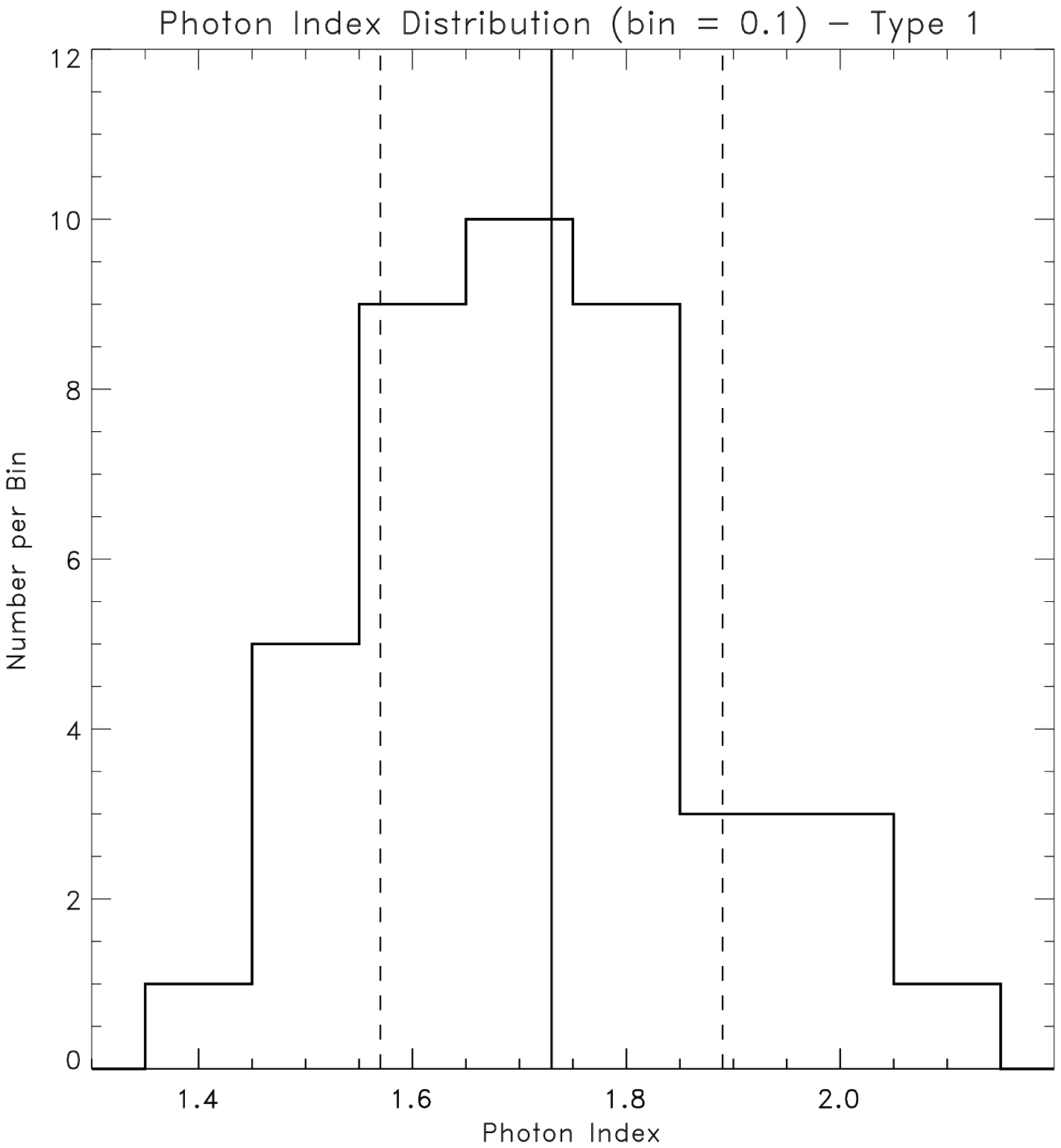}
\caption{Photon index distribution of the entire sample. The dashed lines represent the parameter dispersion area.}
\label{fig1}
\end{figure}

In this letter we report and discuss the general properties of primary continuum parameters, photon index and high energy cut-off, for a large number
of  type 1 AGN.
We calculated their average  values using the  arithmetic mean and study their distribution using the standard deviation\footnote{Lower limits have 
not been considered in this evaluation.}.
We choose to adopt this simple approach because of the  extreme non-gaussianity  of the data points 
and their asymmetric errors, which can be ignored in our analysis.
In Figure~\ref{fig1} is shown the distribution of photon indices of the entire sample. The solid line at $\Gamma$ = 1.73 represents their mean value,
while the dashed lines indicate the relative spread of 0.17.
This mean value is consistent with that previously found in \citet{molina:2009} ($<\Gamma>$ = 1.72, spread 0.2) and with the mean photon index of 
$<\Gamma>$ = 1.73 (spread 0.45) obtained for 156 radio-quiet, X-ray unobscured AGN analysed in the 2--10\,keV band for the \emph{XMM}-CAIXA
catalogue by \citet{bianchi:2009}. It is worth noting that the XMM-CAIXA is the catalogue most compatible with our sample as it contains bright AGN similar to ours,
further it must be emphasised that their mean photon index value has been obtained assuming the presence of the reflection component,
as done in our work. 
It is clear from the figure that the majority of sources fall in the range 1.4--2.1 and that the mean photon index is slightly flatter than 1.9 which is the standard value generally assumed for 
type 1 AGN.

\begin{figure}
\epsscale{.80}
\plotone{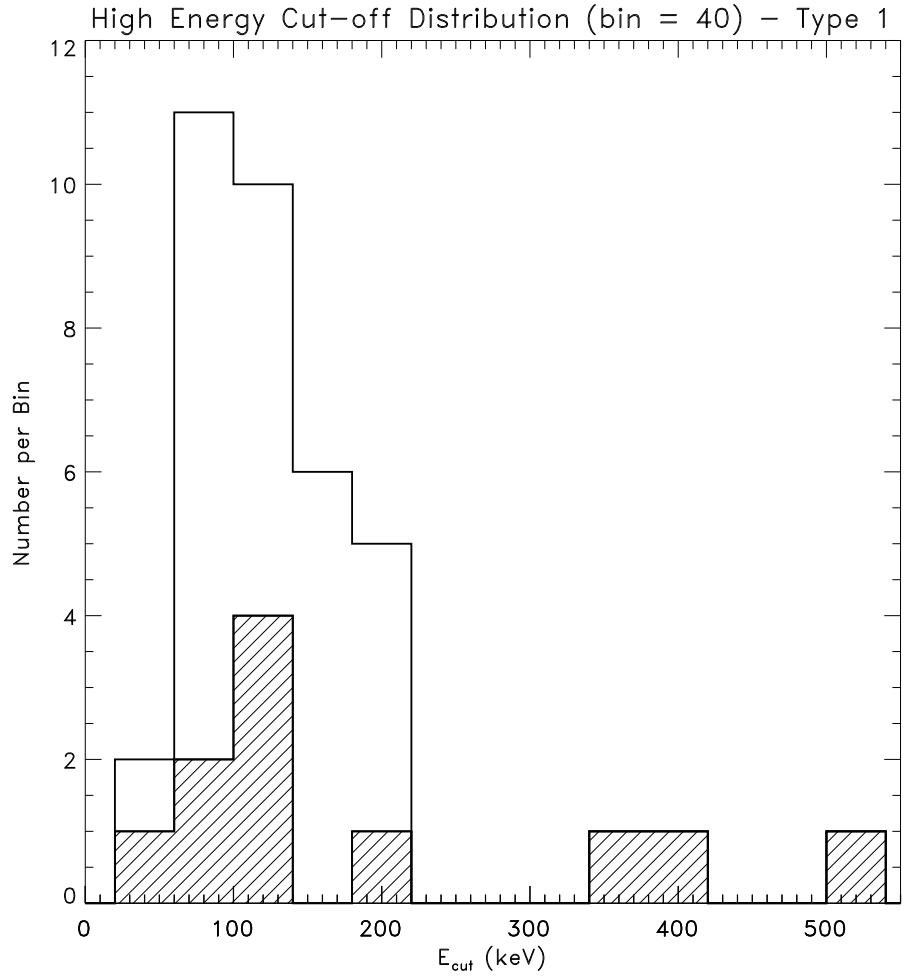}
\caption{High energy cut-off  distribution of the entire sample. 
The diagonally hatched histogram represents sources for which only lower limits of E$_{c}$ are available.}
\label{fig2}
\end{figure}

Of the 41 type 1 AGN analysed in this work, we have been able to constrain the cut-off energy in 26 of them, 
which corresponds to $\sim$63\% of the entire sample.  
In four sources (B3 0309+411B, NGC4593, 2E 1739.1--1210 and IGR J18249--3243) the cut-off energy is found
at energies much higher than the \emph{IBIS/BAT} bandpass and so could not be estimated; for the remaining 11 sources only lower limits are available.   
Figure~\ref{fig2} represents the main result of the present work, as it shows the distributions of E$_{\rm c}$. 
The mean value of the cut-off energy is  128\,keV with a spread of 46\,keV and
its distribution ranges from 50\,keV to 200\,keV, confirming the previous results of \citet{molina:2009}.
Also the majority of the lower limits are below 300\,keV with only a few exceptions (IGR J13109--5552, IGR J17488--3253 and MCG--02--58--022). 
The few lower limits found above 300 keV indicate that higher cut-off energies may be present but only in a small number of AGN.
 However, as we only have data up to 100 keV these lower limits resulting from the fits must be treated with some caution.
It is clear that the cut-off energy in type 1 AGN has a mean value lower than previously found (e.g. \citealt{dadina:2008}) and 
more in line with the first results coming from NuSTAR (see Matt 2013\footnote{http://astro.u-strasbg.fr/~goosmann/gAstronomy\_BH\_Accretion\/Session\_1/session1\_talk\_matt.pdf} and \citealt{brenneman:2013}).

\begin{figure}
\epsscale{.80}
\plotone{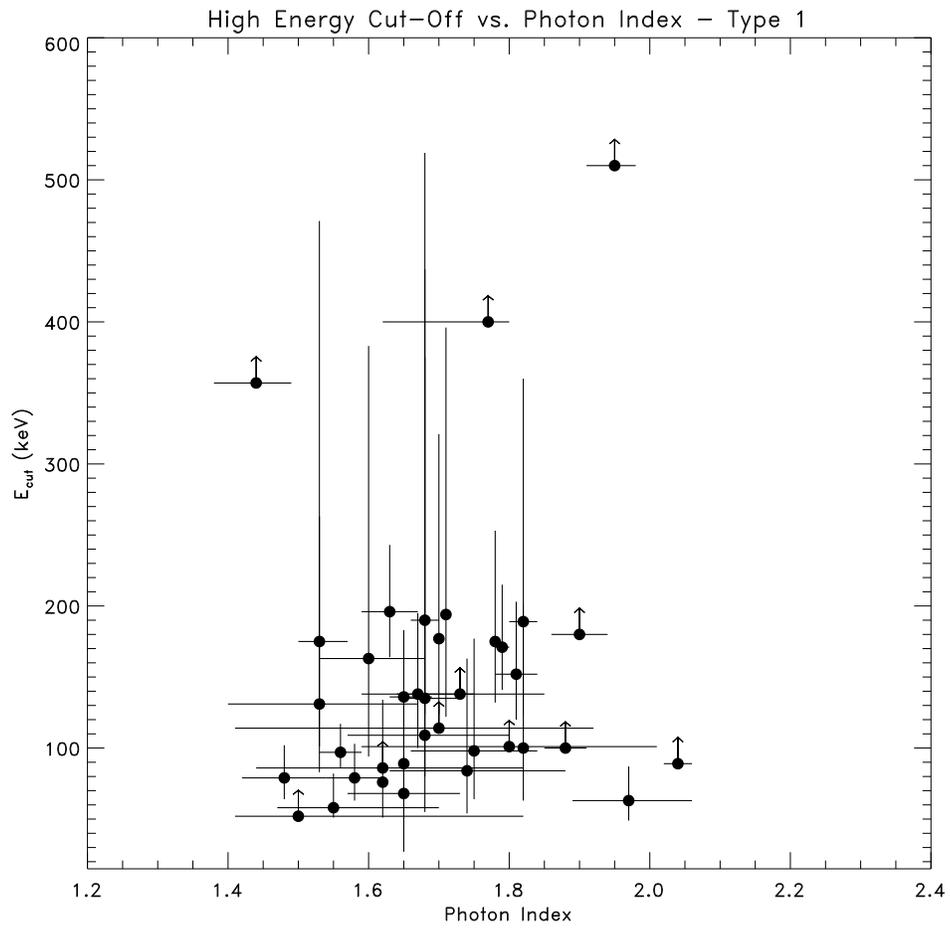}
\caption{High energy cut-off vs. photon index.}
\label{fig3}
\end{figure}

We can use the results on photon indices and cut-off energies to test a possible correlation between them. 
In the past  a trend of  E$_{\rm c}$ increasing  with  $\Gamma$ steepening has been  found \citep{matt:2001, petrucci:2001},
but it is still debated since it is well known that a degeneracy exists between the
photon index and the high energy cut-off  in the spectral model employed.
The high energy cut-offs measured for our sample  are plotted against their respective  photon indices in
figure~\ref{fig3}, but no evident trend is found between these two quantities.
This has also been confirmed by using the Pearson statistical test on the two sets of data. The test  returns   a low
correlation coefficient of \emph{r}$\sim$0.16\footnote{The square of the correlation coefficient \emph{r} is 
normally used as a measure of the association between two variables.} 
(if upper limits are ignored \emph{r} is 0.12); we point out that the errors on the parameters are not considered in this test.
The lack of any correlation between the primary continuum parameters indirectly tests the results of our data analysis confirming that 
the parameter degeneracy does not affect our fitting procedure.

Finally, in order to show the goodness of our fit, in Figure 4 we  plot of the cross-calibration
constants between XMM and BAT (\emph{C$_{\rm BAT}$}) versus XMM and IBIS (\emph{C$_{\rm IBIS}$}).
The good match between the two cross-calibration constants is quite clear, despite the fact we would have expected  some variations between 
\emph{C$_{\rm BAT}$} and \emph{C$_{\rm IBIS}$}, being both datasets averaged over different observation time.
This confirms that long term variability is not so common at high energies in type 1 AGN \citep{beckmann:2007}.
We found mean values of 0.97 with spread of 0.31 and 1.07 with spread of 0.33 
for \emph{C$_{\rm BAT}$} and \emph{C$_{\rm IBIS}$}, respectively. 

\section{Discussion}

In the present work we have been able to determine the main parameters of the primary continuum: $\Gamma$ and E$_{\rm c}$.
Following \citet{petrucci:2001}, within this scenario, we can obtain 
the actual physical parameters of the Comptonizing region from these spectral components.
The plasma temperature kT$_{\rm e}$ is estimated as kT$_{\rm e}$=E$_{\rm c}$/2,
when the optical depth $\tau$$\lesssim$1, while for $\tau$$\gg$1 
kT$_{\rm e}$=E$_{\rm c}$/3 would be more correct. 
Using the following relation from \citet{petrucci:2001}: 
$$
\Gamma-1\simeq\Bigg\{\frac{9}{4}+\frac{m_ec^2}{kT_e\tau(1+\tau/3)}\Bigg\}^{1/2}-\frac{3}{2}
$$
and knowing the temperature, we can calculate the optical depth assuming the spectral index derived from the \texttt{PEXRAV} fit.
We have previously estimated that the mean value of E$_{\rm c}$ for our  sources  is $\sim$130\,keV, ranging
 from 50 to 200\,keV; note that higher cut-off energies may be present but only in a small number of AGN, as also indicated by some of the lower limits found. 
Taking into account the most likely range of E$_{\rm c}$ estimated for our sample, we have derived the most probable 
range of plasma temperatures kT$_{\rm e}$ which is from 20 to 100 keV (or 2 - 12 $\times$10$^{8}$\,K). 
Assuming our average value of $\Gamma$=1.73 and solving the equation for both low and high values of $\tau$ and E$_{\rm c}$,
we get  acceptable solutions for $\tau$ in the range 1 to 4. 
These results are in good agreement with those previously found by \citet{petrucci:2001} for a small sample of Seyfert 1 galaxies and with those found by \citet{molina:2009}
and indicate that the plasma has a typical temperature of (50$\pm$30)\,keV and  an optical depth of $\tau$$<$4.

Our findings on  the high energy-cut-off are consistent with the synthesis models of the CXB which often assume an upper 
limit of $\sim$200\,keV. This assumption is essentially determined by the intensity and shape of the CXB spectrum above 20-30\,keV, which cannot be exceeded.
In fact if one uses  a value of E$_{\rm c}$ of  300\,keV it  becomes difficult to accommodate all available observations and CXB measurements \citep{gilli:2007}.\\

\section{Conclusions}
In this work we presented the broad-band spectral analysis of 41 type 1 AGN of the \emph{INTEGRAL} 
complete sample by fitting together \emph{XMM}, \emph{Swift/BAT} and \emph{INTEGRAL/IBIS} data in the 0.3--100\,keV energy band.
We found that the mean photon index is 1.73 (standard deviation of 0.17) confirming previous results from \emph{XMM} and \emph{INTEGRAL}.
The main result of this work is that for the first time we provide the high energy cut-off distribution for a large  sample of type 1 AGN:
26 objects out of 41 analysed which corresponds to  63\% of sample.
We found a  mean value of $<$ E$_{\rm c}$ $>$  of 128\,keV with a spread of 46\,keV indicating that the primary continuum typically decays at much lower energies 
that previously thought.
We note that this mean value is in line with the  synthesis models of the cosmic diffuse background  which often assume an upper 
limit of $\sim$200\,keV and emphasize  that some of the same of the cut-offs measurements reported here are now being confirmed by \emph{NuSTAR}
\citep{brenneman:2013}. 
It is worth noting that NuSTAR will be hugely advantageous for this science with its high S/N, but only for sources with a relatively low cutoff, i.e., $\lesssim$ 150 keV or so. 
The precision of the measurement of a rollover above this energy would be compromised due to NuSTAR's lack of effective area above 79 keV.
From the primary continuum parameters we have indirectly  estimated the plasma conditions surrounding the black holes, which in our sample has typically a  temperature in a range 20-100\,keV and an optical depth $\tau$$<$4.
A more direct estimate of the electron plasma temperature and corona optical depth in AGN will be provided in the near future by 
\emph{NuSTAR} observations, which cover a broad band from 3--79 keV with much higher sensitivity than  \emph{INTEGRAL/Swift}.
Indeed \emph{NuSTAR} will be able to constrain 
all AGN spectral parameters ($\Gamma$, E$_{\rm c}$ and R) contemporaneously  with great accuracy as has recently been  demonstrated 
 \citep{natalucci:2013}.

\acknowledgments

We acknowledge the Italian Space Agency (ASI) financial programmatic support via contract I/033/10/0


\begin{figure}
\epsscale{.80}
\plotone{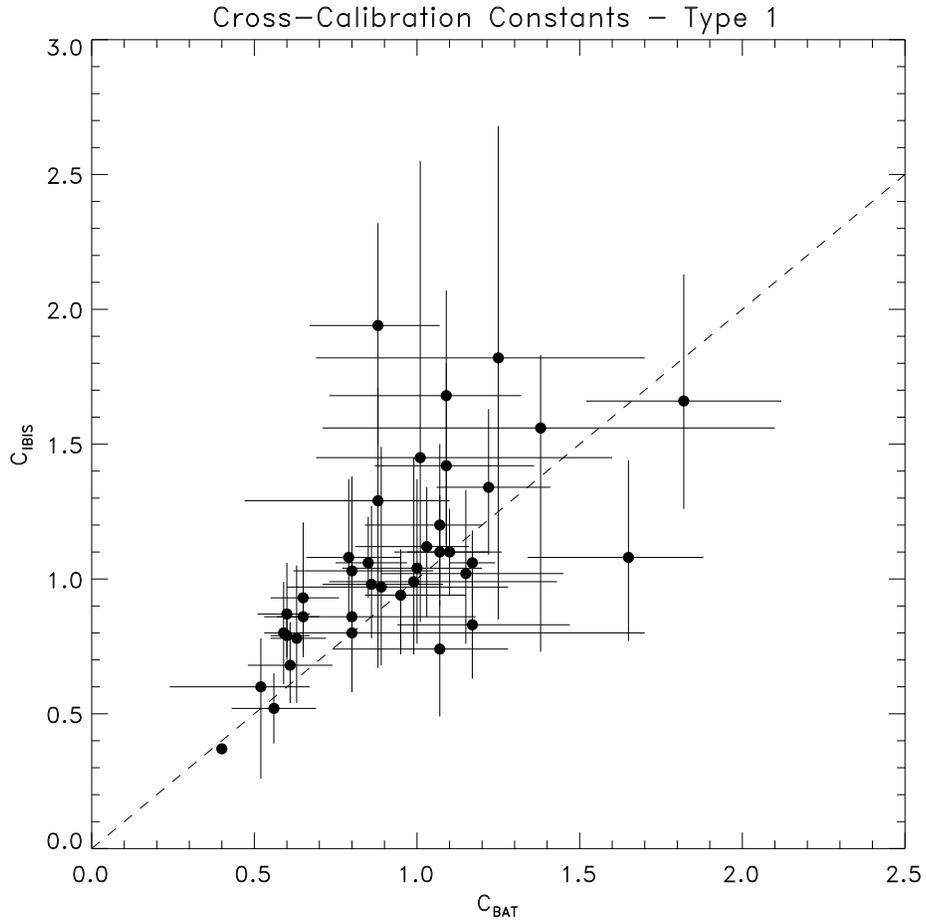}
\caption{Plot of the cross-calibration constants between X-ray and gamma ray data.
    \emph{C$_{BAT}$} is the \emph{XMM} and BAT
    cross-calibration constant and \emph{C$_{IBIS}$} is the
    \emph{XMM} and \emph{INTEGRAL} one. The 1 to 1
    (\emph{C$_{BAT}$}=\emph{C$_{IBIS}$}) line for the constants is
        also shown. }
\label{fig4}
\end{figure}
\clearpage

\clearpage

\begin{deluxetable}{rccccrrrrr}
\rotate
\tabletypesize{\scriptsize}
\tablecolumns{6}
\tablewidth{0pc}
\tablecaption{Main spectral parameters}
\tablehead{
\colhead{Source} &\colhead{kT}               & \colhead{N$_{H}^{FC (a)}$}  & \colhead{N$_{H}^{PC (b)}$ (cf)}   & \colhead{N$_{H}^{W (c)}$ (log $\xi$ - cf)}  & \colhead{$\Gamma$}       & \colhead{E$_{c}$}    & \colhead{C$_{BAT}$}   & \colhead{C$_{IBIS}$}    & \colhead{$\chi^2$ (dof)}\\
                              &    keV                        & ($\times$ 10$^{22}$ cm$^{-2}$) & ($\times$ 10$^{22}$ cm$^{-2}$)  & ($\times$ 10$^{22}$ cm$^{-2}$)        &  &  & &  & 
}
\startdata
IGR J0033+6122   & \nodata                     & 1.66$^{+0.35}_{-0.30}$         & \nodata                                          & \nodata                                             & 1.50$^{+0.32}_{-0.09}$    & $>$52                         & 0.88$^{+0.22}_{-0.41}$  & 1.29$^{+0.42}_{-0.62}$  & 388.3 (365) \\
QSO B0241+62     & \nodata                    & 0.14$^{+0.06}_{-0.05}$          & \nodata                                          & \nodata                                             &1.73$^{+0.12}_{-0.09}$     & $>$138                      & 1.15$^{+0.30}_{-0.26}$   & 1.02$^{+0.31}_{-0.26}$  & 394.4 (492) \\
B3 0309+411         &  \nodata                   & \nodata                                    & \nodata                                         & \nodata                                             &  1.83$^{+0.03}_{-0.04}$             & \nodata                       & 1.64$^{+0.80}_{-0.56}$    & \nodata & 421 (424) \\
3C 111                   & \nodata                    & 0.38$^{+0.01}_{-0.01}$           & \nodata                                          & \nodata                                            &  1.65$^{+0.04}_{-0.02}$  & 136$^{+47}_{-29}$     & 0.60$^{+0.07}_{-0.09}$   & 0.87$^{+0.19}_{-0.17}$  & 1069.6 (1028) \\
LEDA 168563        & 0.37$^{+0.01}_{-0.02}$  & \nodata                             & \nodata                                          & \nodata                                            & 1.70(fixed)                        & 177$^{+144}_{-53}$        & 0.59$^{+0.06}_{-0.06}$    & 0.80$^{+0.19}_{-0.19}$  & 1495 (1327) \\  
4U0517+17            & \nodata                   &  0.10$^{+0.03}_{-0.04}$           & \nodata              & 65.0$^{+34}_{-3.5}$  (2.5 fixed - 0.24$\pm$0.12)    & 1.78(fixed)                        & 175$^{+78}_{-43}$         & 1.10$^{+0.12}_{-0.13}$    & 1.10$^{+0.16}_{-0.16}$  & 1800.3 (1634) \\   
MCG+08-11-011   &  \nodata                  & \nodata                                      & \nodata                                         & \nodata                                            &  1.79$^{+0.01}_{-0.01}$    & 171$^{+44}_{-30}$         & 0.56$^{+0.13}_{-0.13}$    & 0.52$^{+0.13}_{-0.13}$  & 1364.7 (1402) \\ 
Mrk 6                      & \nodata                  &  \nodata                                     & 1.13$^{+0.15}_{-0.17}$ (0.84$\pm$0.02)   & \nodata                             &  1.53$^{+0.14}_{-0.13}$    & 131$^{+132}_{-48}$       & 1.00$^{+0.20}_{-0.23}$    & 1.04$^{+0.33}_{-0.28}$  & 1276.7 (1283) \\
                               &                               &                                                   & 11.4$^{+11.8}_{-8.2}$ (0.13 $\pm$0.10)     &                                          &                                           &                                        &                                           &                                        &                        \\ 
IGR J07597-3842  & 0.16$^{+0.03}_{-0.02}$ & \nodata                              & \nodata                                         & \nodata                                            &  1.58$^{+0.04}_{-0.04}$    &   79$^{+24}_{-16}$         & 0.86$^{+0.22}_{-0.15}$    & 0.98$^{+0.29}_{-0.20}$  & 286.7 (282) \\ 
ESO 209-12           & 0.11$^{+0.03}_{-0.02}$ & \nodata                             & \nodata                                          & \nodata                                            &  1.68$^{+0.05}_{-0.03}$    & 135$^{+302}_{-55}$       & 1.09$^{+0.23}_{-0.36}$    & 1.68$^{+0.39}_{-0.58}$  & 597.1 (534) \\
FRL 1146               & 0.15$^{+0.01}_{-0.01}$  & 0.32$^{+0.06}_{-0.06}$   &  \nodata                                          & \nodata                                           &  1.74$^{+0.14}_{-0.11}$    &  84$^{+79}_{-30}$          & 0.89$^{+0.39}_{-0.29}$    & 0.97$^{+0.52}_{-0.29}$   & 421 (386) \\
SWIFT J0917.2-6221 & 0.09$^{+0.01}_{-0.05}$ & \nodata                         & 2.74$^{+0.20}_{-0.19}$ (0.82$\pm$0.02)  & \nodata                               & 1.65$^{+0.08}_{-0.08}$     & 68$^{+41}_{-41}$         & 1.82$^{+0.30}_{-0.30}$      & 1.66$^{+0.47}_{-0.40}$   & 759 (759) \\
SWIFT J1038.8-4942 & \nodata             & \nodata                                     & 7.11$^{+4.93}_{-6.17}$  (0.68$^{+0.16}_{-0.43}$)     & \nodata               &  1.70$^{+0.22}_{-0.29}$    & $>$114                         & 1.25$^{+0.45}_{-0.57}$      &  1.82$^{+0.86}_{-0.97}$  & 186.8 (178) \\
                                    &                          &                                                  & 2.43$^{+9.48}_{-1.28}$ (0.80$^{+0.01}_{-0.23}$)      &                            &                                           &                                       &                                            &                                         &   \\
NGC 3783               &   \nodata              & 1.16$^{+0.37}_{-0.37}$            &   \nodata                                         & \nodata                                             &  1.75$^{+0.09}_{-0.09}$    & 98$^{+79}_{-34}$          & 0.65$^{+0.11}_{-0.10}$   &  0.93$^{+0.28}_{-0.22}$  & 1605.3 (1768) \\
NGC 4151               &  \nodata               &                                                  & 6.58$^{+0.33}_{-0.33}$ (0.87$\pm$0.01)       & \nodata                             &  1.63$^{+0.04}_{-0.04}$    & 196$^{+47}_{-32 }$       & 0.40$^{+0.01}_{-0.01}$   &  0.37$^{+0.01}_{-0.01}$  & 1580.4 (1449) \\
Mrk 50                    & 2.24$^{+1.00}_{-0.48}$ & \nodata                            & \nodata                                           &  \nodata                                            &   2.04$^{+0.02}_{-0.02}$     & $>$89                           & 1.38$^{+0.72}_{-0.67}$   & 1.56$^{+0.27}_{-0.83}$   & 753.5 (763) \\
NGC 4593              & \nodata                 &   \nodata                                   &  \nodata                                           &  \nodata                                           &  1.82   $^{+0.03}_{-0.03}$     & \nodata                      & 0.60$^{+0.07}_{-0.05}$   & 0.79$^{+0.10}_{-0.08}$   & 1569.9 (1507) \\
IGR J12415-5750  & \nodata                 &   \nodata                                   &  \nodata                                           &  \nodata                                           &  1.53$^{+0.04}_{-0.03}$    & 175$^{+296}_{-74 }$    & 1.07$^{+0.21}_{-0.33}$   & 0.74$^{+0.18}_{-0.25}$   & 455.3 (434) \\
IGR J13109-5552  & \nodata                 &   \nodata                                   &  \nodata                                           &  \nodata                                           &  1.44$^{+0.05}_{-0.06}$     & $>$ 357                        & 1.17$^{+0.30}_{-0.23}$   & 0.83$^{+0.21}_{-0.20}$   & 300.5 (318) \\
MCG-06-30-15       & \nodata                 & 0.38$^{+0.11}_{-0.12}$           &  \nodata                                           & \nodata                                            &  1.97$^{+0.09}_{-0.08}$    & 63$^{+24}_{-14}$          & 1.22$^{+0.19}_{-0.16}$  &  1.34$^{+0.29}_{-0.25}$   & 1112.7 (1038) \\
4U 1344-60            & \nodata                 & 0.93$^{+0.15}_{-0.18}$           & 4.74$^{+1.69}_{-1.56}$ (0.53$^{+0.08}_{-0.10}$)  & \nodata                    &  1.80$^{+0.21}_{-0.21}$   & $>$101                          & 0.61$^{+0.13}_{-0.12}$  &  0.68$^{+0.16}_{-0.14}$  &  644.3 (604) \\
                               &                              &                                                 & 48.22$^{+31.88}_{-20.75}$  (0.34$^{+0.12}_{-0.15}$)                               &                                          &                                        &                                        &                                         &               \\  
IC 4329A                &   \nodata              & 0.35$^{+0.01}_{-0.01}$           & \nodata                                            &  \nodata                                           & 1.81$^{+0.03}_{-0.03}$    & 152$^{+51}_{-32}$        & 0.85$^{+0.12}_{-0.10}$  &  1.06$^{+0.17}_{-0.13}$  &  1699.4 (1617)\\
ESO 511-G030      & 0.27$\pm$0.01    & \nodata                                    &  \nodata                                           & \nodata                                            &  1.82 (fixed)                      & 100$^{+101}_{-37}$      & 0.79$^{+0.16}_{-0.13}$  &  1.08$^{+0.29}_{-0.24}$  &  1119.1 (1049) \\
IGR J16119-6036  & \nodata                &  \nodata                                   &  \nodata                                           &  \nodata                                           &  1.88$^{+0.03}_{-0.03}$    & $>$100                         & 1.09$^{+0.27}_{-0.22}$  & 1.42$^{+0.38}_{-0.29}$  &  598.3 (630) \\
IGR J16482-3036  & \nodata                &  0.07$^{+0.02}_{-0.02}$          & \nodata                                            & \nodata                                            &  1.60$^{+0.08}_{-0.07}$    & 163$^{+220}_{-69}$     & 0.52$^{+0.15}_{-0.28}$ & 0.60$^{+0.18}_{-0.34}$ & 598.3 (576) \\
IGR J16558-5203  & 0.59$^{+0.07}_{-0.06}$  & \nodata                         &  \nodata                                           & \nodata                                            &  1.71 (fixed)                       & 194$^{+202}_{-72}$    & 0.99$^{+0.44}_{-0.26}$  & 0.99$^{+0.46}_{-0.27}$   & 503.38 (550) \\
GRS 1734-292      & 0.18$^{+0.13}_{-0.07}$   & 1.98$^{+0.30}_{-0.22}$ & \nodata                                          & \nodata                                            & 1.55$^{+0.15}_{-0.08}$     &  58$^{+24}_{-7}$         & 1.03$^{+0.13}_{-0.22}$  & 1.12$^{+0.22}_{-0.26}$   & 455.1 (462) \\
2E 1739.1-1210     & \nodata                & 0.17$^{+0.01}_{-0.01}$            &  \nodata                                          & \nodata                                            & 2.12$^{+0.05}_{-0.05}$     &  \nodata                      & 1.07$^{+0.19}_{-0.14}$   & 1.10$^{+0.21}_{-0.17}$   & 602.0 (613) \\
IGR J17488-3253  & \nodata                & 0.37$^{+0.08}_{-0.11}$            &  \nodata                                          & \nodata                                            & 1.77$^{+0.03}_{-0.15}$    & $>$400                       & 0.88$^{+0.19}_{-0.21}$   & 1.94$^{+0.38}_{-0.44}$   & 210.7 (201) \\
IGR J18027-1455   & \nodata               & \nodata                                    &  0.38$^{+0.16}_{-0.11}$ ($<$0.88) & \nodata                                            &1.62$^{+0.20}_{-0.18}$    & $>$86     & 0.95$^{+0.72}_{-0.28}$   & \nodata                           & 538.2 (517) \\
IGR J18249-3243   & \nodata               & \nodata                                    & 6.84$^{+5.69}_{-4.28}$ (0.21$^{+0.11}_{-0.14}$) &  \nodata                    & 2.04$^{+0.04}_{-0.03}$   & \nodata                        & \nodata                            & 0.74$^{+0.27}_{-0.21}$   & 672.7 (734) \\
IGR J18259-0706$^{(d)}$  & \nodata  & 0.90$^{+0.11}_{-0.08}$             & \nodata                                            & \nodata                                           &  1.68$^{+0.12}_{-0.11}$   & 109$^{+410}_{-54}$   & 0.80$^{+0.38}_{-0.27}$   & 0.86$^{+0.42}_{-0.28}$   & 941.2 (979) \\   
3C 390.3                  & 0.69$\pm$0.04  & \nodata                                    & \nodata                                            & \nodata                                           &  1.56$^{+0.03}_{-0.03}$    & 97$^{+20}_{-11}$       & 1.17$^{+0.07}_{-0.07}$    & 1.06$^{+0.12}_{-0.14}$   & 1423.9 (1507)\\
2E 1853.7+1534      & \nodata              & \nodata                                    & 0.23$^{+0.05}_{-0.02}$ ($>$0.85)    & \nodata                                          &  1.65 (fixed)                       & 89$^{+50}_{-26}$      & 0.80$^{+0.25}_{-0.18}$   & 1.03$^{+0.35}_{-0.24}$   & 365.6 (434) \\
NGC 6814               & \nodata               & \nodata                                   & \nodata                                             & \nodata                                          &  1.68$^{+0.02}_{-0.02}$    & 190$^{+185}_{-66}$   & 1.07$^{+0.13}_{-0.23}$   & 1.20$^{+0.30}_{-0.30}$   & 771.0 (770) \\
4C 74.26                 & \nodata               & \nodata                                   &  \nodata               &  0.19 (fixed) (2.6 fixed)                                             & 1.82$^{+0.02}_{-0.02}$    & 189$^{+171}_{-66}$   & 0.63$^{+0.09}_{-0.08}$   & 0.78$^{+0.27}_{-0.24}$   & 1285.5 (1267) \\
S5 2116+81             & \nodata               & \nodata                                   & \nodata                                             & \nodata                                          &  1.90$^{+0.04}_{-0.04}$    & $>$180                       & 1.01$^{+0.59}_{-0.32}$   & 1.45$^{+1.10}_{-0.61}$   & 451.9 (445) \\
IGR J21247+5058    & \nodata              & \nodata                                   & 0.77$^{+0.18}_{-0.13}$ (0.89$^{+0.10}_{-0.06}$  & \nodata                      &  1.48$^{+0.06}_{-0.06}$     & 79$^{+23}_{-15}$       & 0.65$^{+0.05}_{-0.08}$  & 0.86$^{+0.08}_{-0.12}$   & 2276.7 (2555) \\ 
                                 &                           &                                                 & 7.78$^{+2.02}_{-1.66}$ (0.27$^{+0.04}_{-0.05}$                                      &                                            &                                    &                                         &                                         &      \\ 
RX J2135.9+4728   & \nodata              & 0.27$^{+0.01}_{-0.01}$           & \nodata                                             & \nodata                                           &  1.62 (fixed)                       & 76$^{+58}_{-25}$       & 0.80$^{+0.90}_{-0.16}$   & 0.80$^{+0.24}_{-0.19}$  & 712.9 (689) \\   
MR 2251-178           & \nodata             & \nodata   & \nodata                                 & 0.73$^{+0.30}_{-0.17}$  (0.33$^{+0.49}_{-0.22}$ - 0.53$\pm$0.5) &  1.67$^{+0.08}_{-0.08}$    & 138$^{+57}_{-38}$     & 0.95$^{+0.20}_{-0.11}$   & 0.94$^{+0.17}_{-0.22}$   & 990.1 (993) \\
MCG-02-58-022       &\nodata             & \nodata                                     & 4.16$^{+2.60}_{-1.40}$ (0.30$\pm$0.03)   & \nodata                                &  1.95$^{+0.03}_{-0.04}$    & $>$510                      & 1.65$^{+0.23}_{-0.31}$   & 1.08$^{+0.36}_{-0.31}$   & 468.0 (450) \\
\enddata
\tablecomments{Error quotes at 90\% confidence level for one interesting parameter;
(a): N$_{H}^{FC}$ refers to the cold absorption fully covering the source. 
(b): N$_{H}^{PC}$ refers to the cold absorption(s) partially covering the nucleus, with cf being the covering fraction;
(c): N$_{H}^{W}$ refers to the warm absorption, log $\xi$ is the ionization parameter and cf is the covering fraction;
(d): soft excess  fitted with a power law of $\Gamma$=2.5.}
\end{deluxetable}

\clearpage

\end{document}